\documentclass[final,authoryear,3p,times]{elsarticle}

\newcommand{\bibpath}{./} % Path to references
\newcommand{\volume}{\mathop{\ooalign{\hfil$V$\hfil\cr\kern0.08em--\hfil\cr}}\nolimits}
\usepackage{natbib}
\usepackage{amssymb}
\usepackage{graphicx}
\usepackage{verbatim}
\usepackage[mathscr]{eucal}
\usepackage{lineno}
\usepackage{epstopdf} % Inserting .eps-files

\usepackage{hyperref}

\usepackage{multirow} % Multirows and multicolumns in tables
\usepackage{amsmath}  % Typesetting math 
\usepackage{float}    % Fix figures and tables [H]
\usepackage{setspace}
\journal{Appl. Ocean Res. 
{\normalfont Published version is at: 
\url{https://doi.org/10.1016/j.apor.2020.102159}}}

\begin{document}
\begin{frontmatter}

\author{David R. Fuhrman\corref{cor1}\fnref{DTU}}
\ead{drf@mek.dtu.dk}
\cortext[cor1]{Corresponding author}
%%\ead{bjelt@mek.dtu.dk}
%%\cortext[cor1]{Corresponding author}
\author[DTU]{Bjarke Eltard Larsen}

%%\author[METU]{Cuneyt Baykal}
%%\author[DTU]{B. Mutlu Sumer}

%\ead{jf@mek.dtu.dk}

%\title{}
\title{A discussion on ``Numerical computations of resonant sloshing using the modified isoAdvector method and the buoyancy-modified turbulence closure model'' [Appl.~Ocean Res.~(2019), 93, article no.~101829, doi:10.1016.j.apor.2019.05.014]}
  
%%\author[label1,label2]{<author name>}
\address[DTU]{Technical University of Denmark, Department of Mechanical Engineering, DK-2800 Kgs. Lyngby, Denmark}

%\begin{abstract}

%\end{abstract}
%\begin{keyword}
%CFD, turbulence modelling, breaking waves, wave boundary layers
%\end{keyword}
\end{frontmatter}

%\section{Clarifying discussion on the over-production of turbulence beneath waves in two-equation models}
%\label{sec:appendix}
%\linenumbers

First and foremost, the discussors wish to compliment the authors, \cite{Lietal2019}, on their impressive results involving computational fluid dynamics (CFD) simulation of resonant sloshing.
We have written this discussion merely to clarify what we feel are some potential mis-characterizations of our recent work \citep{LarsenFuhrman2018} found within the paper.  While we recognize that these have not been the main emphasis of the discussed paper, we raise the issues below with the simple hope of preventing their further propagation in the literature.  In \cite{Lietal2019} it is stated that ``\cite{LarsenFuhrman2018} proved that classical two-equation RANS closure models are unstable in the two-phase flow. Those closure models were originally developed for single-phase flow, and those turbulence models can be applied to the two-phase flow which is treated as a single continuum mixture.  However, those turbulence models can lead to the overestimation of the turbulence level in the transition region at the [air-water] interface.'' We fear that the statements above mis-characterizes the nature of our work regarding two important issues, as detailed below.

First, the statement above that we proved the (unconditional) instability of several classical two-equation models for two-phase flow is not correct. Building on the prior analysis of \cite{MayerMadsen2000}, the analysis of \cite{LarsenFuhrman2018}, as stated clearly on their p.~424, assumes constant density and hence pertains specifically to the most canonical case of single-phase (and not two-phase) flow.  This restriction in the analysis was merely for the sake of simplicity i.e.~the number of phases being modelled is not of central importance in diagnosing the fundamental cause of the instability (hence over-production) of the turbulent kinetic energy density $k$ and eddy viscosity $\nu_T$ in two-equation turbulence models beneath (non-breaking) surface waves.
%The analysis is hence appropriate for the intended potential flow region beneath propagating (non-breaking) surface waves, which would typically consist of a single phase (water).
The statement above leaves the impression that the instability is somehow linked to models originally designed for single phase conditions being applied in two-phase situations. This is not the case: While our analysis has been made formally assuming constant density, it holds reasonably for the bulk fluid region beneath non-breaking surface waves in one- or two-phase models, since this region (in either case) is effectively comprised of a single (water) phase. This assertion was confirmed with two-phase simulations of propagating waves by \cite{LarsenFuhrman2018}, e.g.~their Figure 3, which still demonstrated the instability (exponential growth) of turbulent kinetic energy density (and hence the eddy viscosity ) essentially as predicted by their single-phase analysis. Again, the instability proved in this context is inherent in the basic traditional turbulence model equations, and does not somehow depend on the number of phases being considered.

Second, the statements above gives the impression that the over-production of turbulence is a problem specific to the air-water interface.
The perception that the over-production of turbulence in two equation
turbulence models beneath surface waves is necessarily linked to the
air-water interface (or near surface) region seems to persist in (at
least some of) the recent literature \citep[similar characterizations
  can be found in
  e.g.][]{Devolderetal2017,Ahmadetal2019,Kamathetal2019,OudaToorman2019}.
We wish to emphasize that, while such characterizations are not necessarily incorrect, they are certainly incomplete.  Over-production of turbulence can certainly occur due to problems specific to the interface region.
%Problems specifically in this region can be effectively remedied in practice through inclusion of the buoyancy production term, e.g.~as in \cite{Devolderetal2017} as well as in the discussed paper.
At the same time, characterizing the analysis of this problem by \cite{LarsenFuhrman2018} as somehow being specific to the interface or near-surface region, as done in the discussed paper and in several of the references above, is not accurate.
%we fear that they may give a mis-leading and innacurate impression that the problem of over-production of turbulence is exclusively linked to the near surface interface region beneath surface waves, which it is not.
On this issue we therefore offer the following clarifying points:
\begin{enumerate}
\item As shown (conditionally) by \cite{MayerMadsen2000} and
  (unconditionally) by \cite{LarsenFuhrman2018}, most (all that have thus far been analyzed) traditional
  two-equation turbulence closure models are unstable \emph{in the entirety
    of the nearly-potential flow region beneath (non-breaking) surface waves where there is finite strain}.  This
  includes, but is certainly not limited to, the near surface region.
\item Relatedly, it is indeed near the surface where the unstable growth rates will be largest, see e.g.~Eq.~(2.12) of \cite{LarsenFuhrman2018}, which was derived from linear wave theory and reads:
\begin{equation}
  \langle p_0 \rangle =
  \frac{k_w^2 H^2\sigma_w^2}{2} \frac{\cosh(2k_wz)}{\sinh^2(k_wh)},
  \hspace{1cm}
  p_0 = 2S_{ij}S_{ij}
\end{equation}
where $\langle\cdot\rangle$ represents period-averaging, $k_w$ is the wave number, $H$ is the wave height, $\sigma_w$ is the wave angular frequency, $h$ is the water depth, $z$ is the vertical distance from the sea bed, and
\begin{equation}
S_{ij} = \frac{1}{2}\left( \frac{\partial u_i}{x_j} + \frac{\partial u_j}{x_i}    \right)
\end{equation}
is the strain-rate tensor, where $u_i$ are the mean velocities in the Cartesian $x_i$ directions.  This shows that $\langle p_0\rangle$, which largely governs the production of turbulent kinetic energy, grows with $z$, a point which has been similarly made by \cite{Kimetal2019}.  It is thus likewise near the surface where this problem will typically first become evident in CFD simulations of surface waves, to be quickly followed by regions further below.  When viewed in animations (see e.g.~the wave train animations available at: \url{https://doi.org/10.11583/DTU.8180708}), this may appear as turbulence originating from or near the surface spreading throughout the water column, but in actuality it is most likely turbulent kinetic energy in the potential flow core region growing exponentially at a rate which varies vertically. 
    %
%\item There may be other numerical scheme-dependent issues which promote over-poduction of turbulence in the near surface region, and buoyancy modification may help with these specific local issues.  Nevertheless, 
    
  \item Adding a buoyancy production term (so-called buoyancy modification) to the $k$-equation, as done in the discussed paper, creates a local sink in the turbulence and eddy viscosity near the air-water interface.  This may remedy issues local to the interface and likewise remove a local ``triggering'' mechanism further below, thus delaying the onset of instability \citep[as effectively shown by][]{Devolderetal2017}.  This term alone does not result in a formally stable turbulence closure, however.  This is clear from the work of \cite{LarsenFuhrman2018} (their analysis, as well as their Figures 3, 4a, 6a,b, 11, and 12a,b), all of which demonstrate severe over-prediction of turbulence beneath waves throughout the nearly-potential flow core region, or consequential un-physical wave decay, even with the buoyancy production term active.  This is likewise clearly demonstrated in the animations referenced above.  As pointed out by \cite{LarsenFuhrman2018}, this is even clear from the simulations of \cite{Devolderetal2017}, their Figures 5b and 7b, both of which show kinematic eddy viscosities which have evolved to be 100--1000 times larger than the kinematic viscosity of water, after wave train simulations lasting approximately $20$ wave periods.  This problem is likewise seemingly evident in the discussed paper, specifically in the uniformly high eddy viscosity field shown in Figure 34a of \cite{Lietal2019}.  Achieving formal stability requires e.g.~further modification of the eddy viscosity, as suggested by \cite{LarsenFuhrman2018}.  We would encourage the authors to consider making a repeat of their simulation with a formally stabilized closure model, for comparison.

\end{enumerate}

We again congratulate the authors on their paper, and sincerely hope the discussion above helps provide clarity on the specific issues raised.
%what we felt were some potential mis-characterizations made within the discussed paper.

%Jacobsen et al. (2012): Hence, instead of determining the production on the basis of the strain rate, the production is based on the rotation of the velocity field. Therefore, no production takes place in the potential part of the flow. Without this modification, a build-up of turbulence is found similar to that reported by Mayer and Madsen [11].

\section*{Acknowledgment}
The discussors acknowledge support from the Independent Research Fund Denmark project SWASH: Simulating WAve Surf-zone Hydrodynamics and sea bed morphology, Grant no. 8022-00137B.
%All authors additionally acknowledge support from the European Community’s Horizon 2020 Programme through the grant to the budget of the Integrated Infrastructure Initiative HYDRALAB+, Contract no. 654110, in the transnational access project HYBRID. The experimental dataset presented in this paper can be downloaded from 
%https:://dx.doi.org/10.5281/ zenodo.1404709

\bibliography{\bibpath/References} % BibTeX references file

\end{document}